\documentclass{ws-procs9x6}

\def\wb{{\overline W}}

\def\halfs{{\scriptstyle \frac 1 2}}
\def\myphi{\hbox{\my\char'010}}
\def\hypp#1#2#3#4{\mya=#1\myb=#1\advance\myb by1{}_{\the\myb}
\myphi_\the\mya\left[{\omega,#2\atop\phantom{\omega,}#3};#4\right]}
\def\hypg#1#2#3{{}_{p+1}\myphi_{p}
\left[{#1\atop{#2\phantom{w}}};#3\right]}
\def\hypq#1#2#3#4{\mya=#1\myb=#1\advance\myb by1
{}_{\the\myb}\myphi_\the\mya\left[{#2\atop#3};#4\right]}
\font\myi=cmmi10 at 11pt
\font\mybb=msbm10 at 11pt
\newcount\mya\newcount\myb\font\my=cmr10 at 12pt
\font\sf=cmss10
\begin{document}

\title{Recent Developments on Ising and Chiral Potts Model}

\author{Jacques H.~H.\ PERK and Helen AU-YANG}

\address{Department of Physics, \\
Oklahoma State University, \\
Stillwater, OK 74078-3072, USA \\
E-mail: perk@okstate.edu}

\maketitle

\abstracts{
After briefly reviewing selected Ising and chiral Potts model results,
we discuss a number of properties of cyclic hypergeometric functions which
appear naturally in the description of the integrable chiral Potts model
and its three-dimensional generalizations.}


\section{Ising Model and Integrable Chiral Potts Model}

\subsection{Z-Invariant Ising Model}

Baxter's $Z$-invariant Ising model is the prototype integrable lattice model
in statistical mechanics. It is ``exactly solvable" for two reasons, namely
because of a complete parametrization in terms of Yang--Baxter rapidities
but also because of reformulations in terms of free fermions. This does not
mean that the calculation of its pair-correlation or its susceptibility is
a straightforward exercise. A more detailed description of the singularity
structure of the zero-field susceptibility of the square-lattice Ising model
has been obtained only recently.\cite{ONGP}

Both integrability features were exploited in our recent studies of the
pair-correlation function and the wavevector-dependent susceptibility of
Ising models with quasiperiodic coupling constants\cite{AJPquasi,APising}
and of the pentagrid Ising model\cite{APpenta} of Korepin.


\subsection{Integrable Chiral Potts Model}

\begin{figure}[ht]
\centerline{\epsfxsize=3.1in\epsfbox{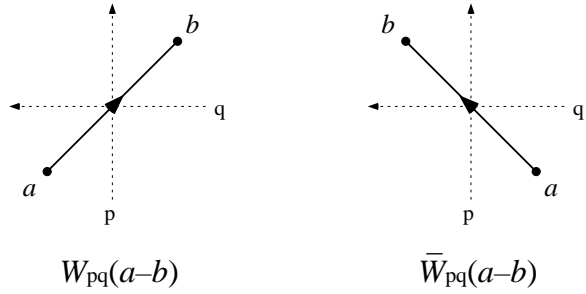}}
\caption{Chiral Potts Model Boltzmann Weights. \label{fig:weights}}
\end{figure}

An $N$-state generalization of the Ising model with fermions replaced
by cyclic parafermions is given by the integrable chiral Potts
model.\cite{AMPTY,BPA,APtani} One version of this model is given in terms
of a square lattice of horizontal and vertical rapidity lines with
rapidities q and p, respectively pointing left and up. After
black-and-white checkerboard coloring of the faces, Potts spins are
placed on the black faces. Boltzmann weights $W(a-b)$ and $\wb(a-b)$ are
assigned to each nearest-neighbor pair of spins in states $\omega^a$ and
$\omega^b$,
\begin{equation}
\omega\equiv{\rm e}^{2\pi{\rm i}/N},
\end{equation}
($a,b=1,\ldots,N$), as in Fig.~\ref{fig:weights}. Here the difference
$a-b$ is to be taken mod$\,N$.
The Boltzmann weights $W$ and $\wb$ can be parametrized as
\begin{equation}
\frac{W_{\rm{\! pq}}(n)}{W_{\rm{\! pq}}(0)}=\!
{\Bigl(\frac{\mu_{\rm p}}{\mu_{\rm q}}\Bigr)}^{\!n}
\prod^{n}_{j=1}\frac{y_{\rm q}\! -\! x_{\rm p}\omega^j}
{ y_{\rm p}\!-\! x_{\rm q}\omega^j},\quad
\frac{\wb_{\rm{\!\! pq}}(n)}{\wb_{\rm{\!\! pq}}(0)}
=\!{(\mu_{\rm p} \mu_{\rm q})}^{\!n}
\prod^{n}_{j=1}\frac{\omega x_{\rm p}\!-\! x_{\rm q}\omega^j}
{ y_{\rm q}\!-\! y_{\rm p}\omega^j}.
\end{equation}
\noindent The rapidities $\rm p$ and $\rm q$ lie on a higher genus curve 
with moduli ${\rm k}$, ${\rm k}'$, with ${\rm k\vphantom{'}}^2+{\rm k'}^2=1$.
The p-curve is parametrized by $(x_{\rm p},y_{\rm p},\mu_{\rm p})$
satisfying the algebraic equations
\begin{equation}
y_{\rm p}^N=(1-{\rm k'}\lambda_{\rm p})/{\rm k},\quad
x_{\rm p}^N=(1-{\rm k'}/\lambda_{\rm p})/{\rm k},\quad
\mu_{\rm p}^N=\lambda_{\rm p}, \label{xymu}
\end{equation}
\begin{equation}
\lambda_{\rm p}+\lambda_{\rm p}^{-1}=
(1+{\rm k'}^2-{\rm k\vphantom{'}}^2{t_p}^N)/{\rm k'},\quad
t_{\rm p}=x_{\rm p}y_{\rm p},
\end{equation}
which follow from the two mod$\,N$ conditions
$W_{\rm{\! pq}}(N)=W_{\rm{\! pq}}(0)$ and
$\wb_{\rm{\! pq}}(N)=\wb_{\rm{\! pq}}(0)$.
Given a value of $t_{\rm p}$ one can choose $|\lambda_{\rm p}|>1$ or
$|\lambda_{\rm p}|<1$. Then $x_{\rm p}$, $y_{\rm p}$, $\mu_{\rm p}$ are
given by (\ref{xymu}) up to powers of $\omega$.


\subsection{Chiral Potts Free Energy and Order Parameters}

Baxter has derived several exact results for the free energy of the
chiral Potts model. Most of his work is based on a set of functional
equations for the transfer matrices.\cite{BBP} An account with results for
all four regimes, with each of $|\lambda_{\rm p}|$ and $|\lambda_{\rm q}|$
$>1$ or $<1$, can be found in Ref.~\refcite{AJPfree}. Baxter also obtained
results for the interfacial tension, which can be much simplified in the
symmetric case.\cite{APrevisit}

For the order parameters of the integrable chiral Potts model we have
\begin{equation}
\langle\sigma_0^n\rangle = {(1-{k'}^2)}^{\beta_n},\quad
\beta_n=\frac{n(N-n)}{2N^2},\quad
(1\le n\le N-1,\quad \sigma_0^N=1),
\end{equation}
which was conjectured\cite{AMPT} early in 1988 and proved only very
recently by Baxter.\cite{Bop1,Bop2}


\section{Cyclic Hypergeometric Functions}
\subsection{Basic Hypergeometric Series at Root of Unity}

The basic hypergeometric hypergeometric series is defined as
\begin{equation}
\hypg{\alpha_1,\cdots,\alpha_{p+1}}
{\phantom{\alpha_1}\beta_1,\cdots,\beta_p}{z}=
\sum_{l=0}^{\infty}{\frac{(\alpha_1;q)_l\cdots(\alpha_{p+1};q)_l}
{(\beta_1;q)_l\cdots(\beta_{p};q)_l(q;q)_l}}\,z^{l},
\end{equation}
where
\begin{equation}
(x;q)_l\equiv\prod_{j=1}^l\;(1-xq^{j-1}),\quad l\ge0.
\end{equation}
Setting first $\alpha_{p+1}=q^{1-N}$ and then
$q\to\omega\equiv {\rm e}^{2\pi{\rm i}/N}$,
we get
\begin{equation}
\hypg{\omega,\alpha_1,\cdots,\alpha_{p}}
{\phantom{\omega,}\beta_1,\cdots,\beta_p}{z}=
\sum_{l=0}^{N-1}
{\frac{(\alpha_1;\omega)_l\cdots(\alpha_{p};\omega)_l}
{(\beta_1;\omega)_l\cdots(\beta_{p};\omega)_l}}\,z^{l}. \label{cychyp}
\end{equation}
We note
\begin{equation}
(x;\omega)_{l+N}=(1-x^N)(x;\omega)_{l}\quad\hbox{and}\quad
(\omega;\omega)_l=0,\quad l\ge N.
\end{equation}
Requiring 
\begin{equation}
{z^N=\prod_{j=1}^p\gamma_j^N,\quad
{\gamma_j}^N=\frac{1-\beta_j^N}{1-\alpha_j^N}}, \label{cyclic}
\end{equation}
we obtain from (\ref{cychyp}) the ``cyclic hypergeometric function"
with summand periodic mod $N$.
Of special importance is the Saalsch\"utz case, defined by
\begin{equation}
z=q=\frac{\beta_1\cdots\beta_p}{\alpha_1\cdots\alpha_{p+1}}
\qquad\hbox{or}\quad
\omega^2\alpha_1\alpha_2\cdots\alpha_{p}=\beta_1\beta_2\cdots\beta_p,
\quad z=\omega. \label{saals}
\end{equation}

The theory of cyclic hypergeometric series is intimately related with
the theory of the integrable chiral Potts model and its generalizations
in three dimensions. We note that our notations differ from those of
Bazhanov and Baxter\cite{BB1,BB2} and of others,\cite{KMS,MSS,SMS,SBMS}
who have an upside-down version of the $q$-Pochhammer symbol $(x;q)_l$.


\subsection{Integrable Chiral Potts Model Weights}

The weights $W$ and $\wb$ of the integrable chiral Potts model can be
written in product form\cite{APrevisit}
\begin{equation}
\frac{W(n)}{W(0)}=\gamma^{n}\,\frac{(\alpha;\omega)_n}{(\beta;\omega)_n},
\qquad\gamma^N=\frac{1-\beta^N}{1-\alpha^N}.
\end{equation}
This is periodic with period $N$.

The dual weights are given by Fourier transform, {\it i.e.}\cite{APfaces}
\begin{equation}
\hat W(k)=\sum_{n=0}^{N-1}\omega^{nk}\,W(n)=
\hypp{1}{\alpha}{\beta\;\;}{\gamma\,\omega^k}\,W(0). \label{fourier}
\end{equation}
They have the same structure as the original weights\cite{APtani,APfaces}
\begin{equation}
\frac{\hat W(n)}{\hat W(0)}=
\hat\gamma^{n}\,\frac{(\hat\alpha;\omega)_n}{(\hat\beta;\omega)_n},\quad
\hbox{with}\quad
\hat\alpha=\gamma,\quad\hat\beta=\frac{\omega\alpha\gamma}{\beta},
\quad\hat\gamma=\frac{\omega}{\beta}. \label{fouriermu}
\end{equation}

\subsection{Summation Formula for ${}_2\myphi_1$\label{susesum}}
The ${}_2\myphi_1$ is exactly summable as a product.\cite{APfaces}
More precisely, we introduce the functions
\begin{equation}
\Delta(z)\equiv(1-z^N)^{1/N},\quad
p(z)\equiv\prod_{j=1}^{N-1}(1-\omega^j z)^{j/N},
\end{equation}
\begin{equation}
p_0(z)\equiv\frac{p(z)}{\displaystyle{\Delta(z)^{(N-1)/2}}}=
\prod_{j=1}^{N-1}\biggl[\frac{(1-\omega^j z)}{\Delta(z)}\biggr]^{j/N},
\label{pzero}
\end{equation}
with all have cuts for $z^N\ge1$ real, with the exception that $p(z)$
is regular on the positive real $z$-axis, where 
$p(1)=\sqrt{N}\Phi_0,\quad\Phi_0\equiv\omega^{(N-1)(N-2)/24}$.

\begin{figure}[ht]
\centerline{\epsfxsize=2.0in\epsfbox{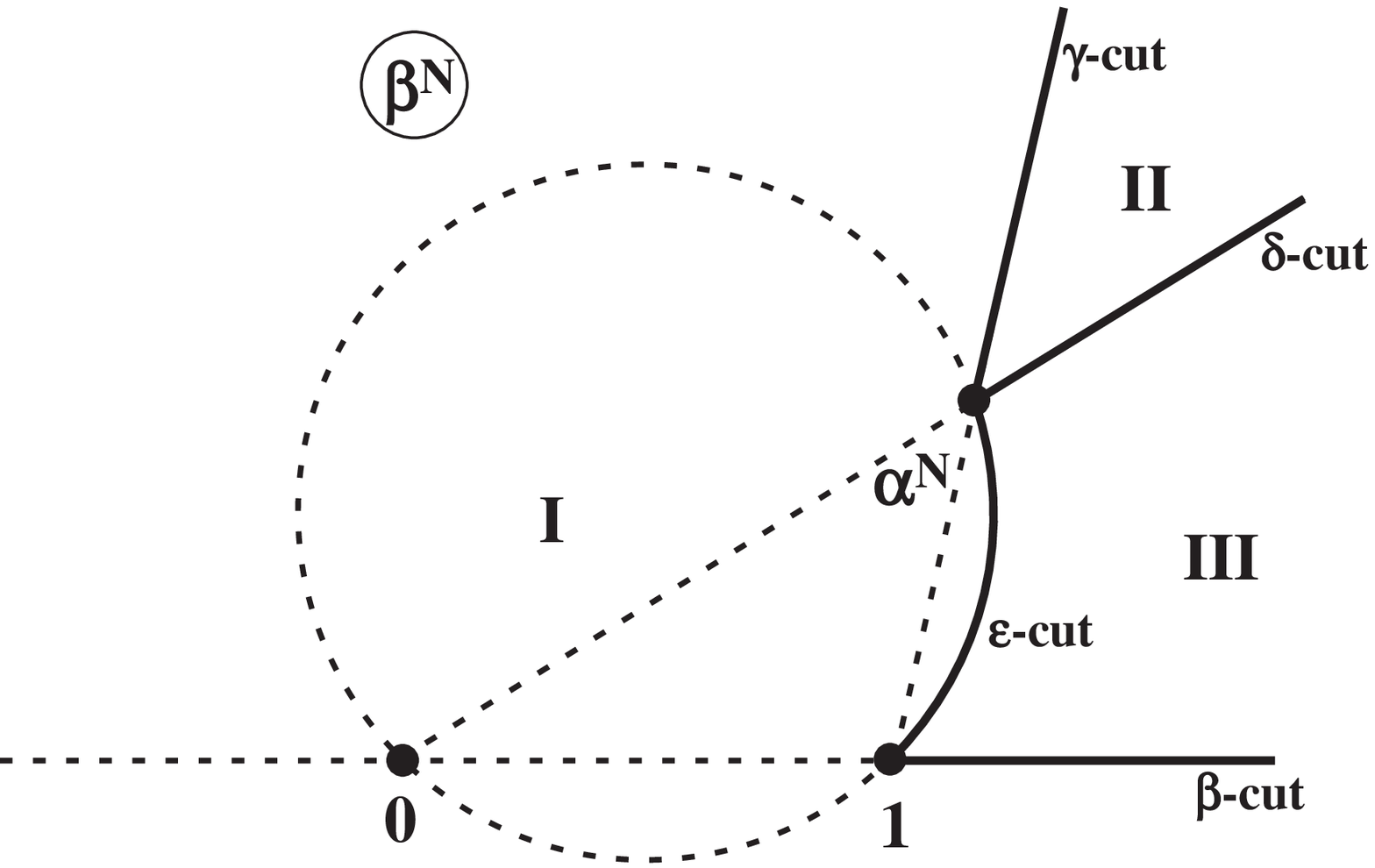}\qquad
\epsfxsize=2.0in\epsfbox{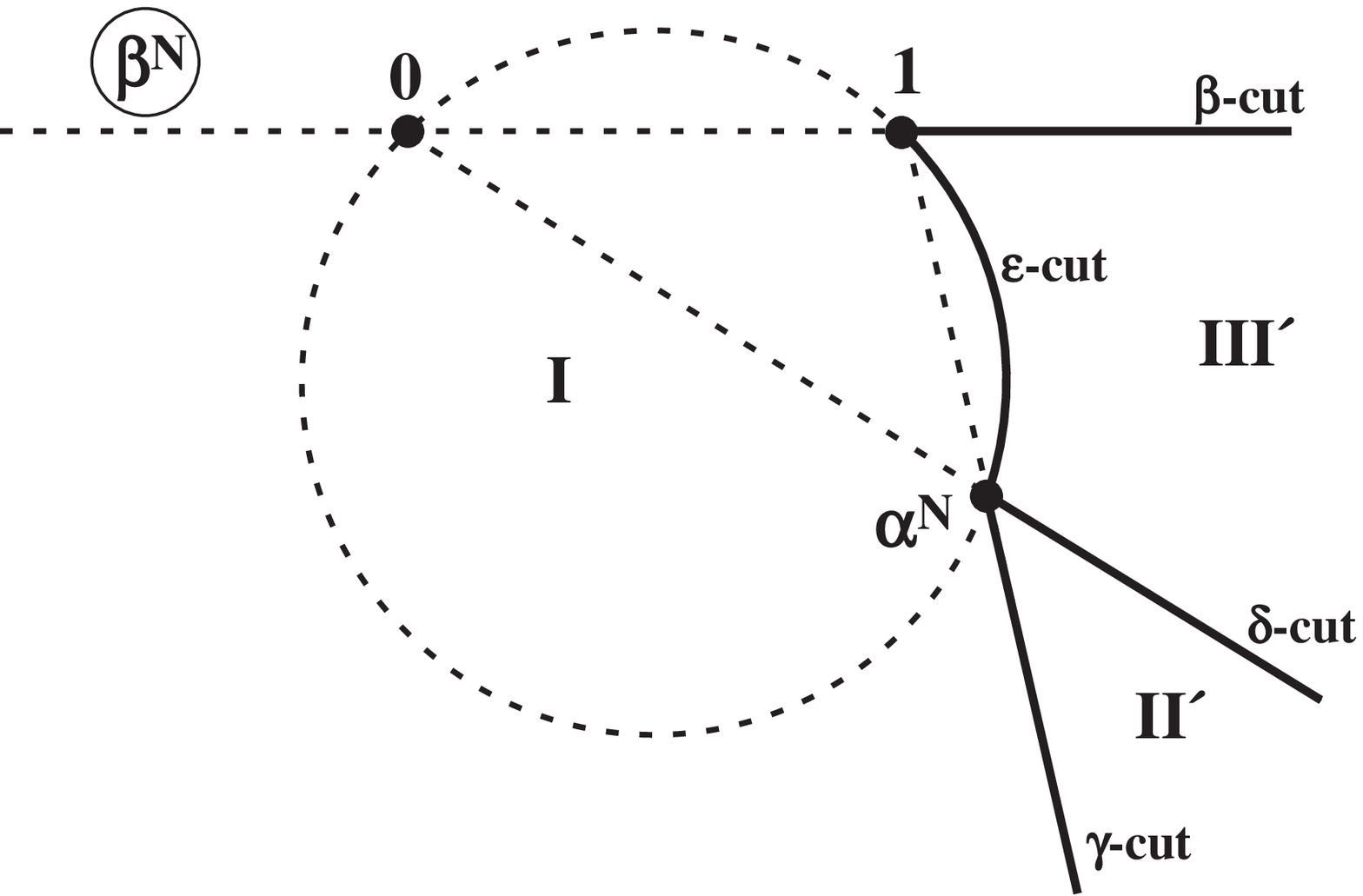}}
\caption{Cut structure in $\beta^N$-plane for Im$\,\alpha^N>0$,
respectively Im$\,\alpha^N<0$. \label{cuts}}
\end{figure}

With these definitions,
\begin{equation}
\hypp{1}{\alpha}{\beta\quad}{{\gamma}}=
F_{\ast}\;\omega^{-\halfs k(k+1)-mk}\,
\frac{N}{\gamma^{\halfs(N-1)}}\,
\frac{p(\beta)p(\gamma)p(\varepsilon)}{p(\alpha)p(1)p(\delta)},
\label{summation}
\end{equation}
where
\begin{equation}
m\equiv\left\lfloor\frac{N}{2\pi}\arg\alpha\right\rfloor,\quad
n\equiv\left\lfloor\frac{N}{2\pi}\arg\beta\right\rfloor,
\end{equation}
with $\lfloor x\rfloor$ the floor of $x$ and
\begin{equation}
\gamma\equiv\omega^k\frac{\Delta(\beta)}{\Delta(\alpha)},\quad
\delta\equiv\frac{\beta}{\alpha},\quad
\varepsilon\equiv\frac{\beta}{\alpha\gamma}.
\end{equation}
The phase factor $F_{\ast}$ can take several values. If we keep $\alpha$
fixed and move $\beta$ in the complex plane, we encounter the cuts in
Fig.~\ref{cuts}.
From a detailed analysis at each cut we find
\begin{eqnarray}
&F_{\,\rm I}=1,\quad
F_{\,\rm II}=\omega^k,\quad
F_{\,\rm III}=\omega^{m-n+k},\quad
&\mbox{if }\;\mbox{Im}\,\alpha^N>0,\nonumber\\
&F_{\,\rm I}=1,\quad
F_{\,{\rm II}'}=\omega^{-k},\quad
F_{\,{\rm III}'}=\omega^{n-m-k},\quad
&\mbox{if }\;\mbox{Im}\,\alpha^N<0.
\end{eqnarray}

Noting
\begin{equation}
(z;\omega)_n\equiv\prod_{j=1}^n\;(1-\omega^{j-1}z),\quad z=0,\ldots,N-1.
\end{equation}
we see that
\begin{equation}
\frac{p(\omega^nz)}{p(z)}=\frac{p_0(\omega^nz)}{p_0(z)}=
\frac{(z;\omega)_n}{\Delta(z)^n}\equiv(\!(z;\omega)\!)_n,
\end{equation}
which is a ``cyclic Pochhammer symbol" $(\!(z;\omega)\!)_{n+N}=
(\!(z;\omega)\!)_n$. On the principal sector $0<\arg z<2\pi/N$, we find
\begin{equation}
p_0(z)p_0(\omega/z)=\omega^{(N^2-1)/12}=
\Phi_0^{\;2}\omega^{(N-1)/4},\qquad
\frac{\Delta(\omega/z)}{\Delta(z)}=\frac{\omega^{n+\halfs}}{z}.
\end{equation}


\subsection{$\hbox{\mybb Z}_4$ Symmetry of ${}_2\myphi_1$}

The Fourier transform (\ref{fouriermu}) defines a transformation $\mu$,
\begin{equation}
\mu:\quad\left\{\begin{matrix}
\alpha&\to&\gamma&\to&\displaystyle\frac{\omega}{\beta}&\to&
\displaystyle\frac{\beta}{\alpha\gamma}&\to&\alpha,\cr\cr
\beta&\to&\displaystyle\frac{\omega\alpha\gamma}{\beta}&\to&
\displaystyle\frac{\omega}{\alpha}&\to&
\displaystyle\frac{\omega}{\gamma}&\to&\beta.\end{matrix}\right.
\end{equation}
From (\ref{fourier}) we may infer 
\begin{equation}
\frac{\hat W(0)}{W(0)}=\hypp{1}{\alpha}{\beta\;\;}{\gamma}.
\end{equation}
Using this and applying Fourier transform $\mu$ four times, 
we find
\begin{eqnarray}
\hypp{1}{\alpha}{\beta\;\;}{\gamma}&=&
\frac{N}{\displaystyle{\hypp{1}{\gamma}{{\!\!\!\!\!\omega\alpha\gamma/\beta}}%
{{\omega/\beta}}}}=
\hypp{1}{{\omega/\beta}}{{\omega/\alpha}}{{\beta/\alpha\gamma}}
\nonumber\\&=&
\frac{N}{\displaystyle{\hypp{1}{{\beta/\alpha\gamma}}%
{{\omega/\gamma}}{\alpha}}}=
\hypp{1}{\alpha}{\beta\;\;}{\gamma},
\end{eqnarray}
which is a $\hbox{\mybb Z}_4$ symmetry.


\subsection{The ${}_3\myphi_2$ identities}

Using the convolution theorem, we find
\begin{eqnarray}
\hypp{2}{\alpha_1,\alpha_2}{\beta_1,\beta_2}{\gamma_1\gamma_2}&\nonumber\\
=N^{-1}&\displaystyle{
\sum_{k=0}^{N-1}\hypp{1}{\alpha_1}{\beta_1}{\omega^{-k}\gamma_1}
\hypp{1}{\alpha_2}{\beta_2}{{\omega^k \gamma_2}},}
\end{eqnarray}
where $\gamma_i=\Delta(\beta_i)/\Delta(\alpha_i)$, $i=1,2$. We can use
the recurrence relation\cite{APtani,APfaces}
\begin{equation}
\frac{\hat W(n)}{\hat W(0)}=
\frac{\displaystyle{\hypp{1}{\alpha}{\beta\;\;}{\gamma\,\omega^n}}}{
\displaystyle{\hypp{1}{\alpha}{\beta\;\;}{\gamma}}}=
\hat\gamma^{n}\,\frac{(\hat\alpha;\omega)_n}{(\hat\beta;\omega)_n}.
\end{equation}
to find
\begin{equation}
\hypp{2}{\alpha_1,\alpha_2}{\beta_1,\beta_2}{\gamma_1\gamma_2}=A\;\;
\hypp{2}{\,\beta_1/\alpha_1\gamma_1,\,\gamma_2}%
{\!\!\!\!\omega/\gamma_1,\,\omega\alpha_2\gamma_2/\beta_2}%
{\omega\alpha_1\beta_2}, \label{convolution}
\end{equation}
with
\begin{equation}
A\equiv N^{-1}\;\hypp{1}{\alpha_1}{\beta_1}{\gamma_1}
\hypp{1}{\alpha_2}{\beta_2}{\gamma_2}.
\end{equation}
More generally, one can generate the symmetry relations of
the cube in the Baxter--Bazhanov model under the 48 elements of the
symmetry group of the cube, see also the work of Sergeev
{\it et al}.\cite{SMS}

The group is generated by two generators. The first one is
$\iota_{\alpha}$: $\alpha_1\leftrightarrow\alpha_2$ resulting in
\begin{equation}
\hypp{2}{\alpha_1,\alpha_2}{\beta_1,\beta_2}{\gamma_1\gamma_2}=
\hypp{2}{\alpha_2,\alpha_1}{\beta_1,\beta_2}{\gamma_3\gamma_4},
\label{iota}
\end{equation}
with $\gamma_3=\Delta(\beta_1)/\Delta(\alpha_2)$ and
$\gamma_4=\Delta(\beta_2)/\Delta(\alpha_1)$, so that
$\gamma_3\gamma_4=\gamma_1\gamma_2$. The second generator is 
$M=\mu^{-1}\otimes\mu$, which results in
\begin{equation}
\hypp{2}{\alpha_1,\alpha_2}{\beta_1,\beta_2}{\gamma_1\gamma_2}=
\frac{N\;\displaystyle{\hypp{2}{\tilde\alpha_2,\tilde\alpha_1}%
{\tilde\beta_1,\tilde\beta_2}%
{\tilde\gamma_1\tilde\gamma_2}}}{
\displaystyle{\hypp{1}{\tilde\alpha_1}{\tilde\beta_1}{\tilde\gamma_1}
\hypp{1}{\tilde\alpha_2}{\tilde\beta_2}{\tilde\gamma_2}}},
\end{equation}
with
\begin{eqnarray}
&&\tilde\alpha_1=\frac{\beta_1}{\alpha_1\gamma_1},\quad
\tilde\beta_1=\frac{\omega}{\gamma_1},\quad
\tilde\gamma_1=\alpha_1,\nonumber\\
&&\tilde\alpha_2=\gamma_2,\quad
\tilde\beta_2=\frac{\omega\alpha_2\gamma_2}{\beta_2},\quad
\tilde\gamma_2=\frac{\omega}{\beta_2}. \label{convinv}
\end{eqnarray}
This is the inverse of (\ref{convolution}). We can use (\ref{summation})
to evaluate the ${}_2\myphi_1$'s, but this will lead to a phase factor
depending on the positions of the $\alpha$'s and $\beta$'s with respect to
the cuts defined by Fig~\ref{cuts}. Eqs.~(\ref{convolution}),
(\ref{iota}) and (\ref{convinv}) are valid in general, independent of
choices of Riemann sheets or branch cuts.


\subsection{Connection with Sergeev, Mangazeev and Stroganov}

In several of the Russian works\cite{KMS,MSS,SMS,SBMS} one uses points,
$p$, $p'$, {\it etc.}, from the Fermat curve $\Gamma$ in homogeneous
notation, {\it i.e.},
\begin{equation}
p\in\Gamma\quad\leftrightarrow\quad p=(x,y,z)\quad\hbox{with}\quad
x^N+y^N=z^N.
\end{equation}
In our affine notation, $p\leftrightarrow\alpha$, $p'\leftrightarrow\beta$,
{\it etc.}, we would identify
\begin{equation}
\alpha\equiv\frac{\omega x}{z},\quad \Delta(\alpha)\equiv\frac{y}{z}=
(1-\alpha^N)\strut^{1/N}.
\end{equation}
The assignment of Riemann sheets and branch cuts is more subtle in their
homogeneous notation. They deal with that by breaking up the curve $\Gamma$
in parts $\Gamma_l^m$,
\begin{eqnarray}
p\in\Gamma_{l;m}\equiv\Gamma_l^m\quad\leftrightarrow\quad
\alpha\equiv\frac{\omega^{m+1}x}{z},\quad
\Delta(\alpha)\equiv\frac{\omega^{-l}y}{z},\nonumber\\ \cr
-\frac{\pi}{N}+l\;\frac{2\pi}{N}<\arg\frac{y}{z}<
+\frac{\pi}{N}+l\;\frac{2\pi}{N},
\quad0<\arg\frac{x}{z}<\frac{2\pi}{N}, \label{gammaml}
\end{eqnarray}
and by using the notation $(p,m)$ for points in $\Gamma_0^m$. How their
notations translate into ours is also indicated in (\ref{gammaml}).

The $\omega$-Pochhammer symbol $(x;\omega)_l$ is defined upside-down and
is not even unique in the various Russian papers. It is to be translated as
\begin{equation}
w(x,y,z|l)\equiv\prod_{s=1}^l\frac{y}{z-x\omega^s}=
\Bigl(\frac{y}{z}\Bigr)^l\frac{1}{(\omega x/z;\omega)_l}
\end{equation}
in Ref.~\refcite{KMS}. However, for
the work of Sergeev {\it et al.}\cite{SMS} one must identify
\begin{equation}
w(p'|m'+\sigma)\equiv\frac{1}{p_0(\omega^{\sigma}\alpha)},\quad
w(p|m+\sigma)\equiv\frac{1}{p_0(\omega^{\sigma}\beta)},
\end{equation}
\begin{equation}
\alpha=\omega^{m'+1}\;\frac{x'}{z'},\quad
\Delta(\alpha)=\frac{y'}{z'},\quad
\beta=\omega^{m+1}\;\frac{x}{z},\quad
\Delta(\beta)=\frac{y}{z},
\end{equation}
with $p_0(z)$ defined in (\ref{pzero}),
as they normalize $\prod_l w(x|l)=1$, not $w(p|0)=1$. Therefore,
for the appendix of Ref.~\refcite{SMS} we have to make the translation
\begin{equation}
{}_r{\hbox{\my\char'011}}_r\left(
{{(p_1,m_1),\ldots,(p_r,m_r)}\atop
{(p'_1,m'_1),\ldots,(p'_r,m'_r)}}\bigg|\;n\right)=
\hbox{\myi C}\;\,{}_{r+1}\myphi_{r}
\left[{{\omega,\alpha_1,\ldots,\alpha_r}\atop
{\beta_1,\ldots,\beta_r\phantom{w}}};z\right],
\label{translate}
\end{equation}
\begin{equation}
\hbox{\myi C}\equiv\frac{1}{\sqrt{N}}\;
\frac{p_0(\alpha_1)\cdots p_0(\alpha_r)}{
p_0(\beta_1)\cdots p_0(\beta_r)},\qquad
z\equiv\omega^n\;\frac{\Delta(\beta_1)\cdots\Delta(\beta_r)}{
\Delta(\alpha_1)\cdots\Delta(\alpha_r)}.
\end{equation}


\subsection{Other Identities for Cyclic Hypergeometric Functions}

One can derive many other identities for the cyclic hypergeometric function
(\ref{cychyp}), (\ref{cyclic}). Without giving explicit expressions, we
list some of the types of identities in Table~1.

\begin{table}[ph]
\tbl{Cyclic hypergeometric identities.}
{\footnotesize
\begin{tabular}{@{}ccc@{}}
\hline
{} &{} &{} \\[-1.5ex]
Conditions & ${}_{p+1}\myphi_p=\prod/\prod$ &
${}_{p+1}\myphi_p\propto{}_{p+1}\myphi_p$\\[1ex]
\hline
{} &{} &{}\\[-1.5ex]
None & ${}_2\myphi_1$ & ${}_3\myphi_2$ \\[1ex]
$z=\omega$ & ${}_3\myphi_2$ & ${}_4\myphi_3$ \\[1ex]
Saalsch\"utz & ${}_4\myphi_3$ & ${}_5\myphi_4$ \\[1ex]
\hline
\end{tabular}\label{table1} }
\vspace*{-13pt}
\end{table}

One type of identity is the evaluation of ${}_{p+1}\myphi_p$
in terms of a ratio of products. This is shown in the middle column
of Table~1. Another type of identity is the proportionality of two
${}_{p+1}\myphi_p$'s where the proportionality factor can be expressed
in terms of ${}_{2}\myphi_1$'s or, equivalently, products. This is
shown in the last column of Table~1. The conditions under which such
identities can be found are listed in the first column.

The two cases where there are no further conditions have been discussed
in previous subsections. Other cases requiring the conditions $z=\omega$
and the more restrictive Saalsch\"utz condition (\ref{saals}) have also
been discussed in Ref.~\refcite{APhyper}. The star--triangle equation
of the integrable chiral Potts model is a special case of the
Saalsch\"utzian ${}_4\Phi_3$ identities.\cite{APfaces,APhyper}

It must be noted that identities of all six types in Table~1 have been
derived by Sergeev, Mangazeev and Stroganov in the appendix of
Ref.~\refcite{SMS}. However, one needs the translation (\ref{translate})
to see the connections with more standard basic hypergeometric notations
and with the Saalsch\"utz condition.

Many other identities can be derived. For example, Watson's
analogue of Whipple's theorem for ${}_8\myphi_7$ reduces to
${}_7\myphi_6\propto{}_4\myphi_3$. Moreover, new identities can be
found in the $N\to\infty$ limit.\cite{APinfty}


\section{Final Remarks}

We have presented several results on the deep connection of the integrable
chiral Potts model with the theory of cyclic hypergeometric functions.
Eq.~(\ref{summation}) with $F_{\ast}$ as specified in Sec.~\ref{susesum} is
new and is easier to use than a formulation with multiple Riemann sheets,
especially when doing numerical computations with it. Finally, translation
(\ref{translate}) is also new and may make the results of Sergeev
{\it et al.}\cite{SMS} more accessible to a wider audience familiar with
basic hypergeometric series.


\section*{Acknowledgments}
It is a pleasure to thank Dr. Molin Ge, Dr. Chengming Bai and the Nankai
Institute of Mathematics for their hospitality and support.

\end{document}